\documentclass[draft]{mn2e}

\title[Sub-mm evolution of Sakurai's Object]
      {The sub-millimetre evolution of V4334~Sgr (Sakurai's Object)}
\author[A. Evans et al.]
      {A. Evans$^{1}$, T. R. Geballe$^{2}$, V. H. Tyne$^{1}$, D.
       Pollacco$^{3}$, S. P. S. Eyres$^{4}$, \newauthor 
       B. Smalley$^{1}$ \\
      $^{1}$Department of Physics, Keele University, Keele,
          Staffordshire, ST5 5BG \\
      $^2$Gemini Observatory, 670 N. A'ohoku Place, Hilo, HI\,96720 \\
      $^3$Department of Pure \& Applied Physics, Queen's University of
          Belfast, Belfast, BT7 1NN \\
      $^4$Centre for Astrophysics, University of Central Lancashire,
          Preston, PR1 2HE
}
\date{Revised version: 22 July}

\pagerange{\pageref{firstpage}--\pageref{lastpage}}

\def\LaTeX{L\kern-.36em\raise.3ex\hbox{a}\kern-.15em
    T\kern-.1667em\lower.7ex\hbox{E}\kern-.125emX}

\newcommand{\Mdot}[2]{\mbox{${#1}\times10^{-{#2}}$\,M$_\odot$~yr$^{-1}$}}
\newcommand{\Msun}{\mbox{\,M$_\odot$}}

\newcommand{\mic}{\mbox{$\,\mu$m}}

\newcommand{\ltsimeq}{\raisebox{-0.6ex}{$\,\stackrel 
	{\raisebox{-.2ex}{$\textstyle <$}}{\sim}\,$}} 
\newcommand{\gtsimeq}{\raisebox{-0.6ex}{$\,\stackrel
	{\raisebox{-.2ex}{$\textstyle >$}}{\sim}\,$}} 

\begin{document}
\label{firstpage}
\maketitle

\begin{abstract}
We report the results of monitoring of V4334~Sgr (Sakurai's Object) at 450\mic\
and 850\mic\ with {\sc scuba} on the James Clerk Maxwell Telescope. The flux density
at both wavelengths has increased dramatically since 2001, and is consistent
with continued cooling of the dust shell in which Sakurai's Object is still
enshrouded, and which still dominates the near-infrared emission.  Assuming
that the dust shell is optically thin at sub-millimetre wavelengths and
optically thick in the near-infrared, the sub-millimetre data imply a
mass-loss rate during 2003 of $\sim$\Mdot{3.4\pm0.2}{5} for a gas-to-dust
ratio of 75. This is consistent with the evidence from $1-5$\mic\
observations that the mass-loss is steadily increasing.
\end{abstract}


\section{Introduction}
V4334~Sgr (Sakurai's Object) continues to be the subject of intense
interest. It is a low-mass ($\ltsimeq0.7$\Msun) star that is retracing
its evolution along the Hertzsprung-Russell diagram following a
very late thermal pulse (Herwig 2001). The star is at the centre
of a faint planetary nebula (PN) of diameter $40''$ (e.g. Pollacco
1999; Bond \& Pollacco 2002) and, after an early incarnation as an
F-type star, it underwent a transformation to a carbon star in 1997.

Little is known observationally about the evolution of the star since
mid-1998, when it was completely obscured by an optically thick dust
shell from which (as of April 2004) it has still not emerged; a
dramatic illustration of the obscuration is given by Bond \& Pollacco
(2002). Since the major dust event of 1998 the $1-5$\mic\ spectum
is black-body-like and is simply that of the dust shell. Analysis
of the post-1998 infrared (IR) observations has shown that the dust
is carbon, primarily in amorphous form (Geballe et al. 2002; Tyne
et al. 2002; Eyres et al., in preparation). The $1-5$\mic\ spectra
during the period 1999-2001 were consistent with a mass-loss that
increased from $\sim\Mdot{1.8}{6}$ to $\sim\Mdot{4.4}{6}$ (for the
gas-to-dust ratio of 75 we adopt here; see \S\ref{75} below), while
the maximum grain radius in a grain size distribution
$n(a)\,da \propto a^{-q}\,da$ ($q\simeq3$) increased by a factor $\sim3$
(Tyne et al. 2002).

We first observed V4334~Sgr at sub-millimetre wavelengths with the
{\sc scuba} instrument (Holland et al. 1999) on the 15-m James Clerk Maxwell
Telescope (JCMT) in 2001 August, and reported marginal detections at
both 450\mic\ and 850\mic\ (Evans et al. 2002, hereafter Paper~I).

We report here the results of our continued monitoring of Sakurai's
Object at these wavelengths.

\section{Observations and data reduction}

\begin{table*}
\centering
\caption{Sub-millimetre flux densities $f_\nu$ of V4334~Sgr. Errors are $1\sigma$.}
\begin{tabular}{ccccccccc} \hline
Date     & MJD        &  \multicolumn{2}{c}{FCF (mJy~mV$^{-1}$)} & $f_\nu$(450\mic) &Detection& $f_\nu$(850\mic)  &Detection& Comment \\ \cline{3-4}
{\small D/M/Y}  & -- 2450000          &  450\mic   & 850\mic                     & (mJy)            &level ($\sigma$) & (mJy)   &level ($\sigma$)& \\ \hline
18/08/2001 & 2140 &  $309\pm72$ & $219\pm30$ & $96.6\pm37.5$  & 3.2  & $10.6\pm3.4$ & 3.5 & Paper~I \\
09/01/2003 & 2649 &  $378\pm49$ & $193\pm~5$ & $55.2\pm14.9$  & 4.2  & $35.7\pm2.4$ & 17 & \\
05/05/2003 & 2765 &  $322\pm~3$ & $202\pm~1$ & $157.6\pm25.8$ & 4.1  & $52.1\pm2.7$ & 20 & \\
15/06/2003 & 2806 &  $440\pm~5$ & $217\pm~2$ & $213.2\pm29.1$ & 7.4  & $61.3\pm2.9$ & 28 & \\
16/06/2003 & 2807 &  $401\pm~2$ & $213\pm~1$ & $113\pm41$     & 2.8  & $58.3\pm3.2$ & 18 & \\ \hline
\end{tabular}
\label{fluxes}
\end{table*}

\begin{figure*}
\setlength{\unitlength}{1cm}
\begin{center}
\leavevmode
\begin{picture}(9.0,5.0)
\put(0.0,4.0){\includegraphics{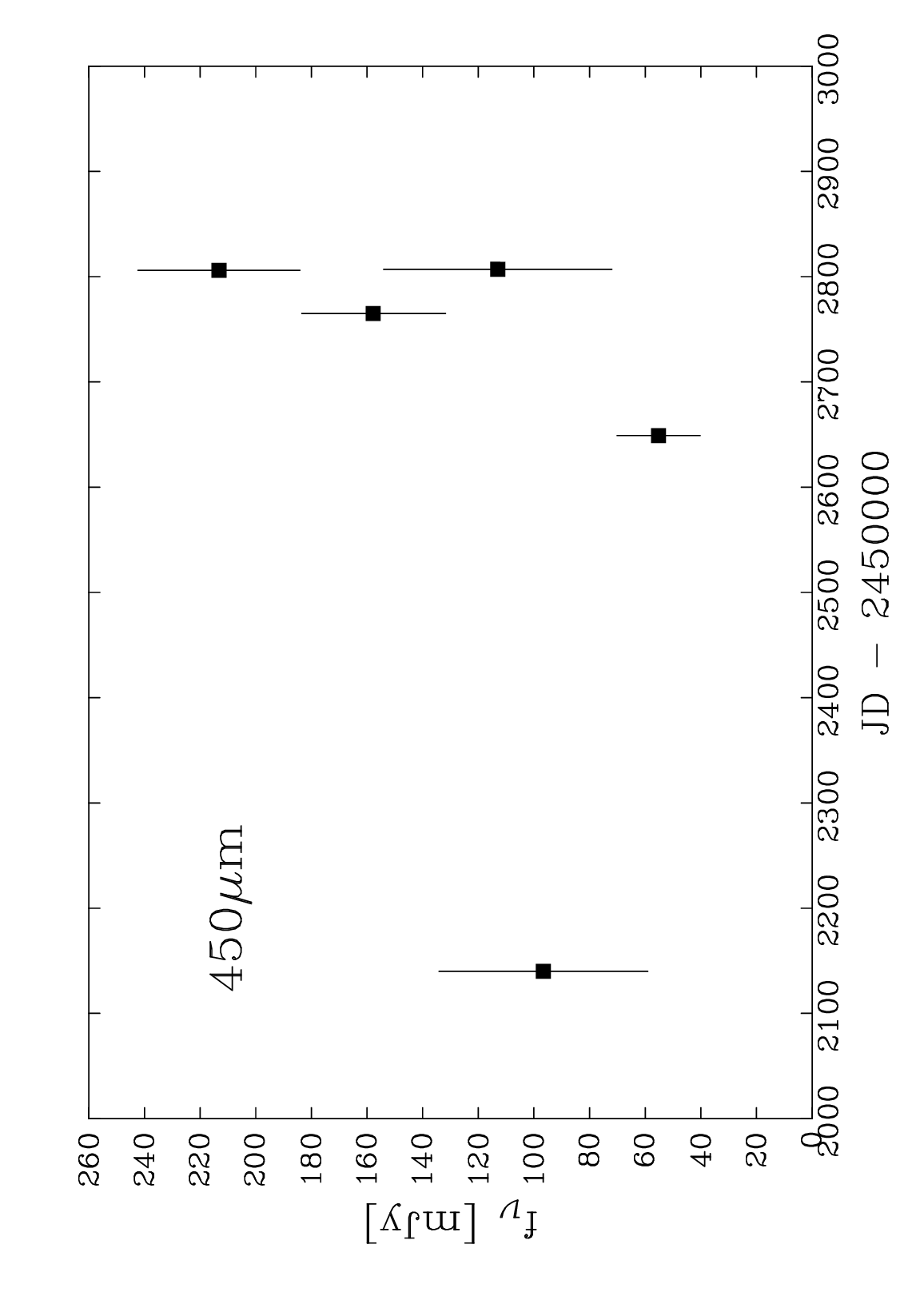}}
\put(0.0,4.0){\includegraphics{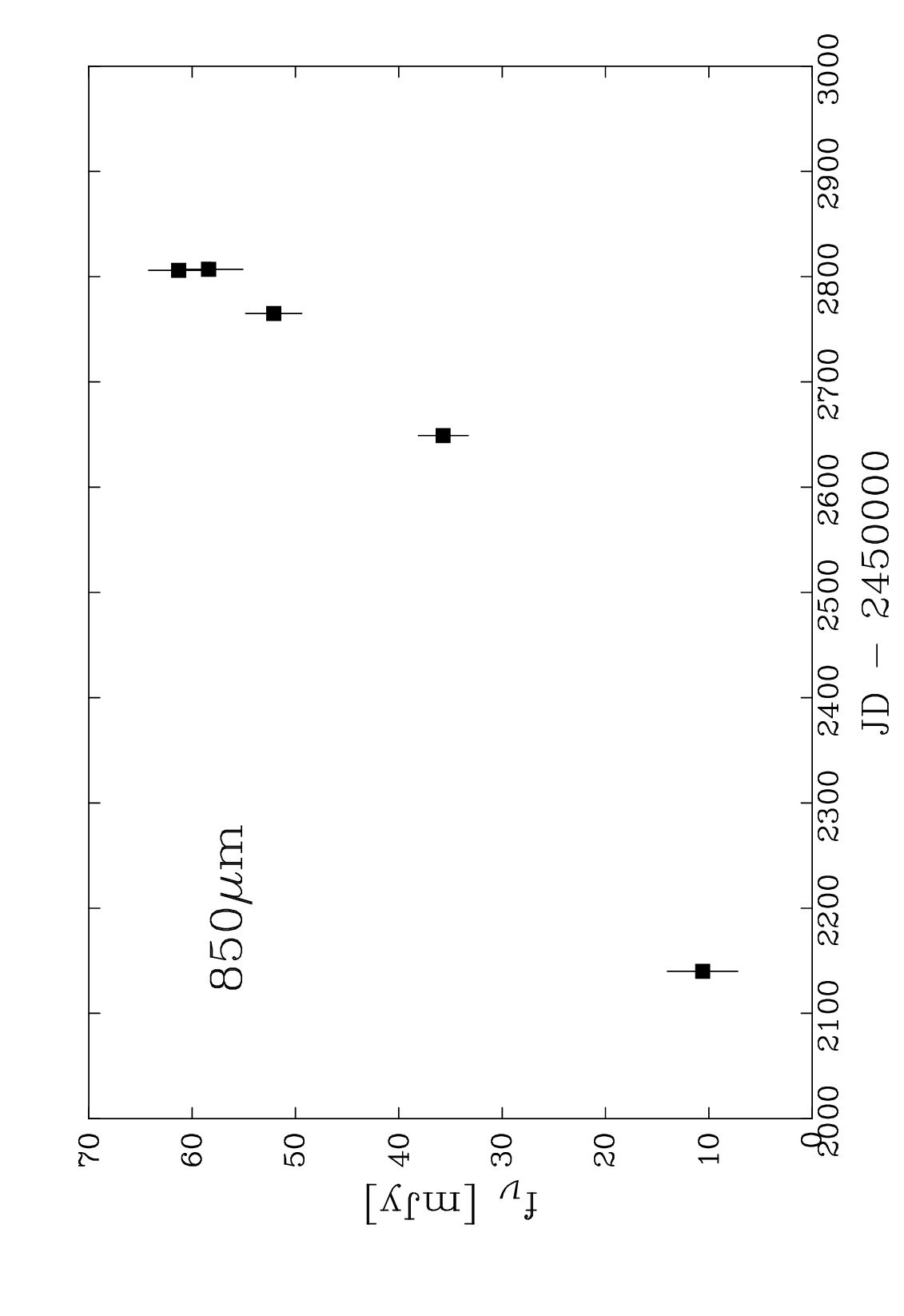}}		 
\end{picture}
\end{center}
\caption[]{The 450\mic\ (left) and 850\mic\ (right) light curves of Sakurai's
Object.}
\label{lc}
\end{figure*}

We have observed Sakurai's Object on the dates given in
Table~\ref{fluxes}, in which we include our 2001 August observation
from Paper~I. On each occasion atmospheric extinction was regularly
monitored by taking skydip measurements, and pointing sources in the
observatory catalogue were used to check telescope pointing and
focussing; Uranus and Mars were used as calibrators. The beamwidths
at 450\mic\ and 850\mic\ are $7''$ and $14''$ respectively.

{\sc scuba} was used in {\sc photometry} mode, which provides simultaneous
observations at 450\mic\ and 850\mic; the observations were carried
out in good (sub-millimetre) weather conditions to ensure that data
at both 450\mic\ and 850\mic\ were obtained. 
However, as is well-known (Holland et al. 1999), radiation at
the shorter wavelength is severely effected by atmospheric water
vapour, and hence detection and calibration is always more
problematic at 450\mic\ even when weather conditions are favourable.
The observations consisted of 2 (or 4) sets of 50 integrations, to
give total on-source integration time of 30 (or 60) minutes. First
order sky subtraction was achieved by chopping $60''$ in azimuth.

The data were reduced using the {\sc surf} package (Jenness \&
Lightfoot 2000). The frames were flat-fielded and corrected for
atmospheric extinction, and the outer bolometers in the array were
used for second order sky subtraction; `spikes' in the data were
clipped to remove any points further than $3\sigma$ from the mean.
The flux conversion factors at 850\mic\ and 450\mic, together with
the detection levels (the signal-to-noise ratios in the measured
signal) for Sakurai's Object are listed in Table~\ref{fluxes}.

Sakurai's Object was strongly detected at both 450\mic\ and 850\mic,
and the flux densities $f_\nu$ are listed in Table~\ref{fluxes}. 
As usual, a significant fraction of the uncertainty in the flux
densities for Sakurai's Object listed in Table~\ref{fluxes} arises
from the uncertainty in the flux calibration, particularly at 450\mic.
In the Table the flux errors include both statistical errors and
calibration uncertainties. The measurement at 450\mic\ for 2003
June 16 is marginal and is not used in what follows.

\section{Discussion}
\label{75}

\begin{table*}
\centering
\caption{Properties of the dust in the environment of Sakurai's Object.
A gas-to-dust ratio of 75 is assumed; see text for details.}
\begin{tabular}{cccccccccc} \hline
Date       & MJD      &  \multicolumn{2}{c}{$M_{\rm d}$\,$T_{\rm d}$
($10^{-5}$\Msun~K)} & $T_{\rm d}$ (K) & \multicolumn{2}{c}{$M_{\rm d}$ ($10^{-7}$\Msun)} && 
                                \multicolumn{2}{c}{$M_{\rm cs}$ ($10^{-5}$\Msun)}\\ \cline{3-4}\cline{6-7}\cline{9-10}
 D/M/Y          & $-2450000$ &  450\mic & 850\mic &   &450\mic & 850\mic && 450\mic & 850\mic\\ \hline
18/08/2001 & 2140  & $4.6\pm1.8$  &  $3.3\pm1.1$ & 504 & $0.9\pm0.4$ & $0.7\pm0.2$ && $0.7$ & $0.5$  \\
09/01/2003 & 2649  & $2.7\pm0.7$  & $11.1\pm0.8$ & 418 & $0.6\pm0.2$ & $2.7\pm0.3$ && $0.5$ & $2.0$  \\
05/05/2003 & 2765  & $7.5\pm1.2$  & $16.2\pm0.8$ & 401 & $1.9\pm0.4$ & $4.0\pm0.5$ && $1.4$ & $3.0$  \\
15/06/2003 & 2806  & $10.2\pm1.4$ & $19.1\pm0.9$ & 395 & $2.6\pm0.4$ & $4.8\pm0.5$ && $1.9$ & $3.6$  \\
16/06/2003 & 2807  &  --         & $18.2\pm1.0$ & 395 & --  & $4.6\pm0.5$  && --    & $3.5$  \\ \hline
\end{tabular}
\label{masses}
\end{table*}

The 450\mic\ and 850\mic\ light curves are shown in Fig.~\ref{lc}.
We see a consistent and significant rise in the 850\mic\ flux
density between 2001 August and 2003 Sepember; the behaviour of
the 450\mic\ data is consistent with the rise at 850\mic, but of
course the 450\mic\ calibration is less certain. In Paper~I we
considered the possibility that the 450/840\mic\ emission
detected in 2001 August might have originated in the old PN but
we concluded that the emission originated in material ejected
in the 1995 event. The fact that the sub-millimetre flux
densities rise so dramatically is consistent with this conclusion.

We first note that the sub-millimetre data may imply a drop in
the $\beta$-index of the dust (defined in the usual way such that
the dust emissivity at frequency $\nu$ is $\propto\nu^\beta$),
from $\sim1.5$ in 2001 August to $\sim0$ in 2003. This may indicate
changes in (a)~the nature of the dust (e.g. as grains grow, or are
processed) and/or (b)~the dust distribution, allowing hotter grains
(which have flatter emissivities, cf. Mennella et al. 1998) to
contribute more to the observed flux. However any conclusion along
these lines must await detailed modelling combining both UKIRT (e.g.
Tyne et al. 2002, Eyres et al., in preparation) and JCMT datasets;
such an analysis will be reported elsewhere.

\begin{figure}
\setlength{\unitlength}{1cm}
\begin{center}
\leavevmode
\begin{picture}(5.0,5.0)
\put(0.0,4.0){\includegraphics{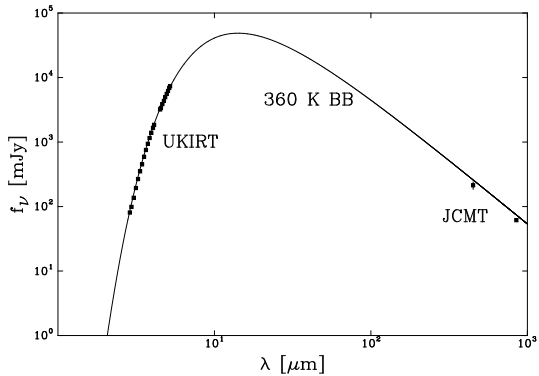}}	 
\end{picture}
\end{center}
\caption[]{Combined UKIRT/JCMT data for Sakurai's Object in late 2003 to
give the 2.5-850\mic\ spectral energy distribution; the UKIRT data have
been binned for clarity. The curve is a 360~K black-body, fitted by eye.
See text for details.}
\label{sak_late03}
\end{figure}

The referee has pointed out that the sub-millimetre rise
could be due to the anticipated rise in effective temperature
as the star retraces its post-AGB evolution towards the PN
phase, and the consequent heating of the inner dust shell.
While observational evidence that this process may have started
has been presented by Kerber et al. (2002), we do not
consider that this accounts for the rise in the sub-millimetre
flux for reasons which may be discerned from Fig.~\ref{sak_late03}.

In this Figure we plot the JCMT data from 15 June 2003
(Table~\ref{fluxes}), together with the 2--5\mic\ data obtained
in September 2003 from our UKIRT programme (Eyres et al., in
preparation); we assume that variations are sufficiently slow
that the data can be combined. Both sets of data are reasonably
fitted by a $\sim360\pm10$~K black-body (cf. Equation~(\ref{temp})
below, and $350\pm30$~K for the UKIRT data alone; Eyres et al. 2004),
suggesting that the 2--5\mic\ and sub-millimetre emission have common
origin.

We might envisage a scenario in which the sub-millimetre
emission in Fig.~\ref{sak_late03} comes from inner, hotter,
dust whose near-IR emission does not penetrate the outer
region of the dust shell, while the observed near-IR
emission comes from dust in the outer shell having very
steep $\beta$-index. However such a situation would be
{\it ad hoc} and contrived. Indeed that the $\beta$-index
is close to zero in late 2003 is supported both by the
450/850 flux ratio, and by the closeness of the 2--850\mic\
data to a black-body at $\sim360$~K (see
Fig.~\ref{sak_late03}).

We therefore make the assumptions that (i)~the emission we detect
at 450\mic\ and 850\mic\ is from the dust shell (any emission at
these wavelengths from the star -- whatever its effective temperature
-- is at the few $\mu$Jy level at most), (ii)~the dust shell is optically
thin at sub-millimetre wavelengths and (iii)~the sub-millimetre
emission is on the Rayleigh-Jeans tail.

We justify (ii)~by noting that
the optical depth $\tau$ at wavelength $\lambda$ (in \mic) is
$\sim 0.55 \, \tau(0.55\mic)/\lambda(\!\mic)\ll1$, since the optical
depth in the visual is $\sim10$ (see Tyne et al. 2002), and (iii)~is
valid if the dust temperature is $\gtsimeq200-300$~K. Eyres et al.
(2004) find that a $1-5$\mic\ spectrum of Sakurai's Object obtained
on 2003 September~8 is well-described by a black-body at temperature
$\sim350\pm30$~K, consistent with continued cooling of the dust shell;
thus (iii)~is certainly justified.
We therefore have that
\begin{eqnarray}
 M_{\rm d} \: T_{\rm d} & = & \frac{f_\nu \, D^2 \, \lambda^2}{2
 \gamma \kappa \, k} \nonumber \\
 \label{eq}
 & \simeq & \frac{A}{\gamma} \left ( \frac{f_\nu}{\mbox{mJy}} \right ) 
    \: \left ( \frac{D}{\mbox{2\,kpc}} \right )^2  \: \Msun\mbox{~K} \:\: ,
\end{eqnarray}
where the constant $A=9.60\times10^{-7}$ for 450\mic, and $6.23\times10^{-6}$
for 850\mic. In Equation~(\ref{eq}) $M_{\rm d}$ is the dust mass, $D$ is the
distance of Sakuai's Object (see Kimeswenger 2002), $T_{\rm d}$ is an
appropriate value for the dust
temperature (see below), $\kappa$ is the absorption coefficient for the dust,
and $\gamma$ depends on physical conditions in the dust shell (e.g.
$\gamma\equiv1$ for an optically thin, isothermal shell).

\begin{figure}
\setlength{\unitlength}{1cm}
\begin{center}
\leavevmode
\begin{picture}(5.0,5.0)
\put(0.0,4.0){\includegraphics{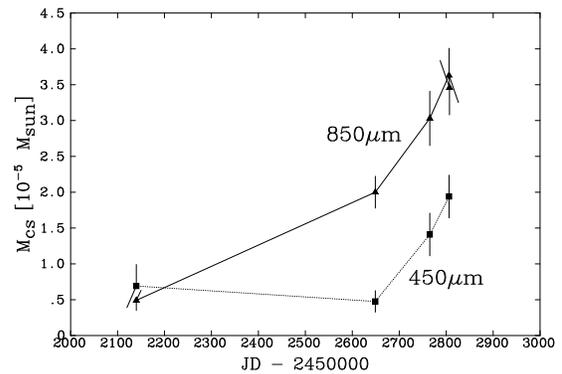}}	 
\end{picture}
\end{center}
\caption[]{Increase in circumstellar mass around Sakurai's Object as
deduced from 450\mic\ data (squares) and 850\mic\ data (triangles), for
a gas-to-dust ratio of 75. Lines are included to guide the eye. See
text for details}
\label{dmass}
\end{figure}

Although Equation~(\ref{eq}) applies to an optically thin,
isothermal shell, it is also accurate, to within a factor
$\simeq 2/(1+(R_{\rm i}/R_{\rm o})^{1/2}) \leq 2$ (where
$R_{\rm i}, R_{\rm o}$ are respectively the inner and outer
radii of the dust shell), to the case of a dust shell with
grain density law $n(r)\propto r^{-2}$ that is optically
thick at short ($\sim$ optical/near-IR) wavelengths,
optically thin in the sub-millimetre, and for which the dust
temperature depends on distance $r$ from the star as
$T_{\rm d}\propto r^{-1/2}$ (see Ivezi\'{c} \&
Elitzur 1997). In this case $T_{\rm d}$ in Equation~(\ref{eq})
is the temperature of the coolest dust; we take this to be the
dust seen in the $1-5$\mic\ data, which still detect the dust
`pseudo\-photosphere' (Tyne et al. 2002, Eyres et
al., in preparation). It is not unreasonable to assume that the dust
shell around Sakurai's Object is geo\-metrically thick, so
that $R_{\rm o}/R_{\rm i}\gg1$; for this case $\gamma=2$
and we use this value here (see also below).

For the distance, we take $D=2$~kpc (see Kimeswenger 2002),
and we have taken $\kappa$ from Mennella et al. (1998; their
$\kappa$ values for amorphous carbon at 295~K, which is likely
to be close to that appropriate for Sakurai's Object).

We give values of $M_{\rm d}$\,$T_{\rm d}$, computed from
Equation~(\ref{eq}), in Table~\ref{masses} for both wavelengths.
For each epoch, we have estimated the temperature $T_{\rm d}$
of the coolest dust by (i)~fitting a black-body to $1-5$\mic\
data obtained in the course of our IR monitoring programme
(see Tyne et al. 2002, Eyres et al. 2004, and paper in
preparation) and (ii)~fitting an exponential decline to the
time-dependence of $T_{\rm d}$. We find:
\begin{equation}
 T_{\rm d} \, (\mbox{K}) \simeq 817 \,\, \exp \left [ \frac{(815.5-t')}{2735.3}
\right ] \: \: ,
\label{temp}
\end{equation}
where $t'=\mbox{MJD}-2450000$ and the uncertainty in $T_{\rm d}$
is typically $\ltsimeq10$\% in reproducing the values in Tyne et
al. (2002) and Eyres et al. (2004). We stress that this expression
is not physically meaningful, and we will refine it as we obtain
further IR data; we use it simply as a means of
interpolating the value of $T_{\rm d}$ at the time of our {\sc scuba}
observations. These values are included in Table~\ref{masses},
together with the corresponding dust masses $M_{\rm d}$; the
uncertainties in $M_{\rm d}$ include an uncertainty of 10\% in
estimating $T_{\rm d}$.

Neither the values of
$M_{\rm d}$\,$T_{\rm d}$, nor of $M_{\rm d}$, should of course
depend on wavelength, and the discrepancies between the 450\mic\
and 850\mic\ values are due to the uncertainties in the 450\mic\
calibration, the error in the interpolated temperature,
uncertainty in the properties of the dust, and the fact that the
$1-5$\mic\ emission differs somewhat from a black-body.
Nevertheless the data at both wavelengths are consistent with a
steady increase in the dust mass between 2001 August and 2003 June.

The dust mass may be converted to a circumstellar mass $M_{\rm cs}$
if we assume a gas-to-dust ratio, and that the gas-to-dust ratio
remains constant in time. The gas-to-dust ratio for solar abundances
and in the interstellar medium is $\simeq120$ (e.g. Whittet 2003),
while a value 200 was assumed by Tyne et al. (2002). These values
are unlikely to be appropriate for the H-deficient, He- and C-rich
wind of Sakurai's Object. If we take the abundances for 1996 July
3 from Asplund et al. (1999), assume that all O is locked up in CO
(see Eyres et al. 2004) and that the remaining C is tied up in dust,
we get a gas-to-dust ratio of 26; this is most likely a lower limit.

We assume an intermediate value of 75 here; the uncertainties
in the gas-to-dust ratio and in the value of $\gamma$ clearly lead
to a corresponding uncertainty in dust mass, but we believe that
{\em changes} in the latter should be reliable. The corresponding
circumstellar masses are given in Table~\ref{masses}; the dependence
of circumstellar mass on time is shown in Fig.~\ref{dmass}.

We use these data to determine the average mass-loss rate over
the period 2001 August to 2003 June. Using the 850\mic\ data, we
find that $\dot{M}\simeq\Mdot{1.6\pm0.3}{5}$ over this period; the
corresponding value from the 450\mic\ data is \Mdot{4.9\pm4.4}{6}.
Taking the 2003 observations only, the 450\mic\ and 850\mic\
data give $\dot{M}\simeq\Mdot{3.3\pm0.3}{5}$ and
$\dot{M}\simeq\Mdot{3.6\pm0.3}{5}$ respectively, with a weighted mean
of $\dot{M}\simeq\Mdot{3.4\pm0.2}{5}$ (cf. Fig.~\ref{dmass}); for
comparison, the maximum mass-loss rate for post-AGB stars is
$\sim\Mdot{1}{4}$ (van Winckel 2003).

Tyne et al. (2002) used the
{\sc dusty} code (Ivezi\'{c} \& Elitzur 1997) to fit the $1-5$\mic\
data with a carbon dust shell and, adjusted for a gas-to-dust ratio of 75,
found that $\dot{M}$ increased from \Mdot{1.8}{6} to \Mdot{4.4}{6}
over the period 1999 May to 2001 September. Our results suggest
that the mass-loss from Sakurai's Object continues to rise.

\section{Conclusions}

We have presented 450\mic\ and 850\mic\ photometry of Sakurai's
Object, and find that the flux densities at both wavelengths have
increased steadily over the period 2001 August to 2003 June. These
data, combined with dust temperatures obtained from out IR monitoring
programme, lead to a mass-loss in 2003 of $\sim$~\Mdot{3.4\pm0.2}{5},
consistent with a continuing rise since our IR observations of 1999-2003.

Monitoring of this object at sub-millimetre wavelengths is continuing
and further observations will be presented elsewhere.

\section*{ACKNOWLEDGMENTS}

We thank the referee, Stefan Kimeswenger, for comments
that helped to clarify some of the discussion.
TRG is supported by the Gemini Observatory, which is
operated by the Association of Universities for Research in Astronomy,
Inc., on behalf of the international Gemini partnership of Argentina,
Australia, Brazil, Canada, Chile, the United Kingdom, and the United
States of America. The JCMT is operated by the Joint Astronomy Centre
on behalf of the Particle Physics and Astroonomy Research Council
(PPARC) of the United Kingdom, the Netherlands Organization for
Scientific Research and the National Research Council of Canada.
We thank the PPARC Panel for the Allocation of Telescope Time for
supporting of this project. Data reduction was carried out
using hardware and software provided by PPARC.

\bsp

\label{lastpage}


\begin{thebibliography}{}
\bibitem{asplund} Asplund M., Lambert D. L., Kipper T., Pollacco D.,
         Shetrone M. D., 1999, A\&A, 343, 507
\bibitem{bond} Bond H. E., Pollacco D., 2002, Keele Workshop on Sakurai's
         Object, Eds A. Evans \& B. Smalley, Ap\&SpSci, 279, 31
\bibitem{evans1} Evans A., Geballe T. R., Tyne V. H., Pollacco D.,
         Eyres S. P. S., Smalley B., 2002, MNRAS, 332, L69 (Paper~I)	    
\bibitem{eyres1} Eyres S. P. S., Smalley B., Geballe T. R., Evans A.,
         Asplund M., Tyne V.H., 1999, MNRAS, 307, L11
\bibitem{eyres2} Eyres S. P. S., Geballe T. R., Tyne V. H., Evans A., Smalley B.,
         Worters H. L., 2004, MNRAS, 350, L9
\bibitem{geballe} Geballe T. R., Evans A., Smalley B., Tyne V. H., Eyres
         S. P. S., 2002, Keele Workshop on
         Sakurai's Object, Eds A. Evans \& B. Smalley, Ap\&SpSci, 279, 39
\bibitem{herwig} Herwig F., 2001, ApJ, 554, L71
\bibitem{scuba} Holland W. S., et al., 1999, MNRAS, 303, 659
\bibitem{dusty} Ivezi\'{c} Z., Elitzur M., 1997, MNRAS,  8287, 799
\bibitem{surf} Jenness T., Lightfoot J. F., 2000, Starlink User Note
         216, Starlink Project, CLRC
\bibitem{kerber} Kerber F., Pirzcal N., de Marco O., Asplund M., Clayton G. C.,
         Rosa M. R., 2002, ApJ, 581, L39
\bibitem{kimeswenger1} Kimeswenger S., 2002, Keele Workshop on
         Sakurai's Object,  Eds A. Evans \& B. Smalley, Ap\&SpSci, 279, 79
\bibitem{mennella} Mennella V., Brucato J. R., Colangeli L., Palumbo P.,
         Rotundi A., Bussoletto E., 1998, ApJ, 496, 1058	   
\bibitem{pollacco} Pollacco D., 1999, MNRAS, 304, 127
\bibitem{tyne2} Tyne V. H., Evans A., Geballe T. R., Eyres S. P. S.,
         Smalley B., D\"{u}rbeck H. W., 2002, 334, 875
\bibitem{araa} van Winckel H., 2003, ARAA, 41, 391
\bibitem{whittet} Whittet D. C. B.,2003, {\it Dust in the Galactic environment},
         2nd edition, Institute of Physics Publishing, Bristol \& Philadelphia
\end{thebibliography}
\end{document}